\documentclass{siamltex}
\usepackage{amsmath,graphicx,subfigure,epsfig}

\newcommand{\bx}{{\bf x}}
\newcommand{\bu}{{\bf u}}
\title{A Multiscale Model of Biofilm as a Senescence-Structured Fluid}
\author{Bruce P. Ayati\thanks{Department of Mathematics, Southern
Methodist University, Dallas, TX 75205 ({\tt ayati@smu.edu}), supported by NSF award DMS-0609854.} \and
  Isaac Klapper\thanks{Department of Mathematical Sciences and Center
    for Biofilm Engineering, Montana State University, Bozeman, MT
    59717 ({\tt klapper@math.montana.edu}),
supported by NIH award 5R01GM67245.}}

\date{}

\begin{document}

\def\trnum{2006-02}
\def\trauths{Bruce P. Ayati and Isaac Klapper}
\def\trtitle{A Multiscale Model of Biofilm as a Senescence-Structured Fluid}
%
%
%
%

\voffset -.25in

\thispagestyle{empty}
\fontsize{12}{12}
\setcounter{page}{0}

\begin{center}
  \centerline{
  \epsfig{figure=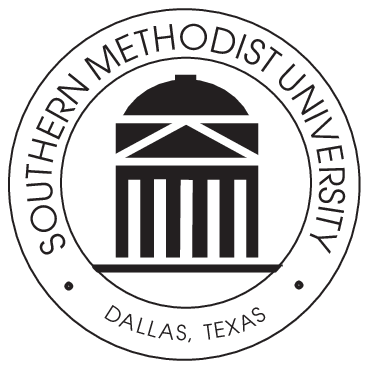,height=1.5in,width=1.5in}
  }

  \vspace{1.25in}

  \begin{minipage}[t]{4in}
    \begin{center}
  
      {\bf \trtitle

      \vspace{24pt}

      \trauths

      \vspace{12pt}

      SMU Math Report \trnum 
      }
    \end{center}
  \end{minipage}

  \vspace{3in}

  {\Large\bf D}{\large\bf EPARTMENT OF} 
  {\Large\bf M}{\large\bf ATHEMATICS} 

  \vspace{3pt}

  {\Large\bf S}{\large\bf OUTHERN} 
  {\Large\bf M}{\large\bf ETHODIST} 
  {\Large\bf U}{\large\bf NIVERSITY} 
\end{center}

\voffset 0in

\normalsize
\newpage

\baselineskip=.33truein
\bigskip

\maketitle
\begin{abstract}
We derive a physiologically structured multiscale model for biofilm
development.  The model has components on two spatial scales, which
induce different time scales into the problem.  The macroscopic
behavior of the system is modeled using growth-induced flow in a
domain with a moving boundary.   Cell-level processes are incorporated
into the model using a so-called physiologically structured variable
to represent cell senescence, which in turn affects cell division and
mortality.  We present computational results for our models which shed
light on modeling the combined role senescence and the biofilm state play in
the defense strategy of bacteria.

\end{abstract}



\section{Introduction}
\label{intro}

In this paper we derive a physiologically structured multiscale model
for biofilm development.  The model has components on two spatial
scales, which induce different time scales into the problem.  The
macroscopic behavior of the system is modeled using growth-induced flow in a
domain with a moving boundary, following
\cite{multispeciesBiofilm,DockeryKlapperSIAP01} .   Cell-level
processes are incorporated into the model using a so-called
physiologically structured variable to represent cell senescence,
which in turn affects cell division and mortality.  We use
``senescence" to mean ``the organic process of growing older and showing
the effects of increasing age"\footnote{www.thefreedictionary.com}.

The multiscale nature of physiologically and spatially structured
population models, such as those in this paper and in
\cite{JMB-proteus,MMStumor06,EnS,FnPnW}, differs from more typical
multiscale systems where the smaller spatial scales have the faster
time scales.   In the structured multiscale systems, the dynamics of
the relevant physiology of individuals within a population are
homogenized to a distribution of a
representative trait, such as age, size or senescence.  Although the
underlying physiological system may have a very fast time scale (such
as the protein network within a cell that controls the cell cycle),
the distribution of the representative trait may evolve relatively
slowly compared to the dynamics in space or in the reaction terms (such as the age distributions
used to represent a tumor cell's position in the cell cycle in
\cite{MMStumor06}, or a {\em Proteus mirabilis} multinuclear filament
cell's length in \cite{JMB-proteus,EnS}).

The derivation of the model in this paper follows that of Alpkvist
and Klapper \cite{multispeciesBiofilm}, with the addition of the
explicit physiological structure in the bacteria populations based on
the notion of bacteria senescence demonstrated in \cite{StewartEtAl2005}
for cells with symmetric division and
\cite{BarkerWalmsley1999,MortimerJohnson1959} for cells with
asymmetric division.  We also include explicit tracking of inert cell
populations, which includes necrotic cells.

To our knowledge, prior to this work,
physiological structure has only been integrated into spatial models
where motion is due to migration or taxis, represented by diffusion
terms in the model equations \cite{JMB-proteus,MMStumor06,EnS,FnPnW}.
Here, instead, motion is driven by growth-induced expansive
stress, a much different mechanism that requires inclusion
of a force balance equation.

This paper is organized as follows.   We first motivate the
problem subject and derive the structured
multiscale model of biofilm growth.  Following this, we present a
nondimensionalization and then a spatially homogeneous steady-state
age distribution, which in combination help to illustrate the
differing age structures occurring in
different places in the biofilm.  Finally, we provide computational
results for our models which shed
light on the combined role senescence, and the self-organization into a biofilm state, play in
the defensive capabilities of bacteria.

\section{Biofilms and Age Dependence}
\label{biofilmDiscussion}

A biofilm is a collection of microorganisms, typically bacteria,
enclosed within a self-secreted polymeric matrix. These films
are generally attached on one side to a solid boundary and,
on the other, access substrates (e.g. oxygen) through a free surface,
Figure \ref{biofilmFig}. See \cite{StoodleyEtAl02} for a review. Biofilm properties change
over long times (weeks) -- it is possible to describe a maturation process
in the development of a particular biofilm. That is, biofilms
demonstrate aging effects. We are thus motivated
to extend basic biofilm models to include age dependence.

In fact, it seems that individual bacteria themselves suffer
age dependence in the form of senescence. This property had been observed
for some time in asymmetric dividers \cite{BarkerWalmsley1999,MortimerJohnson1959}; more recently senescence
has also been noted in the symmetric divider {\it Escherichia coli}
\cite{StewartEtAl2005}. Under normal
conditions, senescent cells make up a small percentage of
the total population. However, aging (over medium time scales) may
provide an effective defense against short time scale
environmental disruptions but without affecting microbial
community vitality during normal conditions. That is, we posit
that multi-time scale behavior allows a powerful defense
mechanism. In a recent paper
\cite{persistenceSenescence}, it was argued
that cell senescence offers a simple explanation for the phenomenon
of bacterial persistence.
Bacteria exhibit the phenomenon
of ``persistence'', the tendency for a small number of cells
within a larger population to tolerate a wide range of antimicrobial
challenges \cite{BalabanEtAl04,Gilbert90,Keren04}.
In particular, the mechanism for this tolerance was suggested
to be that senescent cells were less active and hence less
susceptible than younger, more vigorous ones. Then, once the
antimicrobial attack ceases, the persisting cells could
produce new, vital cells which in turn would be capable
of regenerating the colony. Previously, others
have argued that persisters were phenotypic variants
\cite{BalabanEtAl04, Cogan06, Lewis01, RobertsStewart04, RobertsStewart05}.

Colony defense through persister cells is likely to be
especially effective in biofilms where
surviving persister cells, though perhaps small in number, have
the opportunity to be
protected by the biofilm matrix
\cite{Lewis01}. As a result of this matrix
they may have a particularly
conducive environment for re-population once the antimicrobial challenge
has ended. Thus, in order to demonstrate that senescent cells
can distribute themselves throughout the biofilm and as
a particular application of our age dependent biofilm model,
we compute the spatial and temporal variation of age dependence.
These senescent cells are produced on a medium time-scale
(approximately 1 day): fast relative to the biofilm maturation
time but slow compared to metabolic times. Thus persisters are
generated quickly enough so that they can be an effective defense
mechanism for mature biofilms but not so quickly that they
interfere with competitive fitness.

\section{Derivation of the Model}
\label{model}

\begin{figure}[t]
\begin{center}
\includegraphics[height=1.5in]{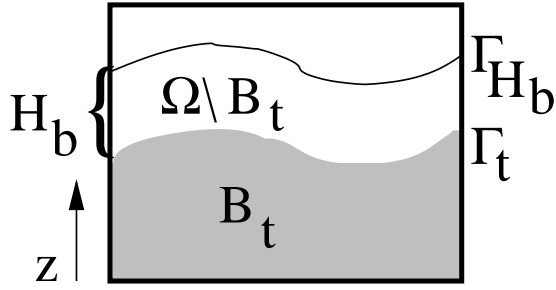}
\caption{Spatial domains for the senescence-structured biofilm model.}
\label{biofilmFig}
\end{center}
\end{figure}

We consider a spatial domain $\Omega$ consisting of stratified subdomains
$B_t$ for biomass and $\Omega \backslash B_t$ for the bulk fluid.
There are two moving interfaces in $\Omega$: $\Gamma_t$ separating
$B_t$ from the rest of $\Omega$, and a bulk-substrate interface
$\Gamma_{H_b}$ that is a fixed height $H_b$ above $\Gamma_t$.  The
biofilm rests on a surface, denoted by a lower boundary, $\Gamma_B$.
The spatial domains are illustrated in Figure \ref{biofilmFig}.

We let $b_i(t,\bx,a)$, $i=1,\ldots,N_b$, denote the densities of the
bacteria phenotypes in time $t \geq 0$, space $\bx \in B_t$, and
senescence $a \geq 0$, and let ${\bf J}_i^b$ denote their respective
fluxes.  The component of $\bx$ representing height is denoted by $z$.
Similarly, $c_i(t,\bx)$, $i=1,\ldots,N_c$, denote the substrate
concentrations in time $t \geq 0$ and space $\bx \in \Omega$,
and ${\bf J}_i^c$ denotes their respective fluxes. In addition
to active cell types, we allow for the presence of inert cells,
including necrotic cells, that do not use or
produce substrates, do not grow, and are not merely in a quiescent state.
Lack of senescence allows us to list these separately from the $b_i$
because these cells do not have any age dependence.
We let $n_i(t,\bx)$ denote inert
cells of type $b_i$  of all ages, and ${\bf J}_i^n$ their
respective fluxes.

Conservation of biomass yields equations
\begin{multline} \label{bs_i}
\frac{\partial b_i}{\partial t} + \frac{\partial
  (g_i(a,c_1,\ldots,c_{N_c}) \, b_i)}{\partial a} + \nabla \cdot {\bf
  J}_i^b \\
=  -\hat{\mu}_i(a,c_1,\ldots,c_{N_c},b_1(t,\bx,\cdot),\ldots,b_{N_b}(t,\bx,\cdot))
  b_i(t,\bx,a) \\ + \hat{f_i}(b_1(t,\bx,a),\ldots,b_{N_b}(t,\bx,a)),
\end{multline}
for $i=1,\ldots,N_b$
where $\hat{\mu}_i$ is the inactivation or ``death'' modulus with dependence on
senescence, substrate concentrations and the densities of all bacteria
phenotypes of all ages, and $\hat{f}_i$ is the rate of net change to
phenotypes $i$ from all other phenotypes.  The terms $\hat{f}_i$
allows the possibility that bacteria have the capability of changing
their phenotype in response to stimuli.  A model that incorporated
change due to mutation would do so in the age boundary condition.  The
senescence rate $g_i$ represents the physical wear-and-tear
experienced by an aging individual in response to nutrient and/or
oxygen uptake and exposure to waste.

Due to the close relationship between senescence and chronological
age, for this paper we make the simplifying assumption that $g_i
\equiv 1$, for $i=1,\ldots,N_b$.  Further below, we
will specify senescence as a function of age for use in both the
inactivation modulus, $\hat{\mu}$, and the fecundity, $\hat{\beta}$, defined
below.  Equations (\ref{bs_i}) then become the age- and space-structured
equations
\begin{multline} \label{b_i}
\frac{\partial b_i}{\partial t} + \frac{\partial b_i}{\partial a} + \nabla \cdot {\bf J}_i^b \\
= -\hat{\mu}_i(a,c_1,\ldots,c_{N_c},b_1(t,\bx,\cdot),\ldots,b_{N_b}(t,\bx,\cdot))
b_i(t,\bx,a) \\ + \hat{f_i}(b_1(t,\bx,a),\ldots,b_{N_b}(t,\bx,a)).
\end{multline}

Observations in \cite{StewartEtAl2005} showed that even in symmetric cell
division, one of the two new cells contains older material and overall
less vitality than the other (referred to as ``old pole" and ``new pole"
cells, resp.).  This results in a physiologically structured
mathematical representation of the bacterial cell cycle that is closer to that
for birth-death processes in animals than what has been commonly used
to represent cell division \cite{MMStumor06, Webb89}.  These models
were built on the assumption that a mother cell divided into two
daughter cells of equal and high vitality, thereby assigning each
daughter cell a senescence of zero and removing the mother cell from
the population.  In our old-pole/new-pole formulation rather, the mother cell
remains in the population and continues to undergo senescence from
the point it had at cell division while giving rise to a single daughter
cell with senescence zero.  This notion of senescence allows two
implications, based on the account in \cite{StewartEtAl2005}, that are
relevant to the model in this paper.   First, old-pole cells grow
slower than new-pole cells produced in the same division. Second, old-pole cells become inert at a higher rate than new-pole cells.

Although a more elaborate model would include explicit size structure
similar to what was done in models in \cite{MMStumor06, Webb89}, we can
take advantage of continuous senescence and time to incorporate the
first old-pole/new-pole issue into a fecundity term,
$\hat{\beta}_i(a,c_1,\ldots,c_{N_c},b_1(t,\bx,\cdot),\ldots,b_{N_b}(t,\bx,\cdot))$,
with dependence on age, substrate concentrations and the densities of all
bacteria phenotypes of all ages.  Differences in individual sizes
influence volume fractions, represented by giving mother cells with
larger offspring a corresponding higher fecundity.   The fecundities,
$\hat{\beta}$, account not just for differences in daughter size due
to the mother's size, but also heterogeneities in the mean growth
rates across phenotypes $i$.  A third more minor property of cell senescence
mentioned in \cite{StewartEtAl2005}, that
new-pole cells are {\em marginally} more likely to divide sooner than
old-pole cells, can also be included in $\hat{\beta}$. (Similarly, 
we can incorporate the second property of higher
incidence of becoming inert into $\hat{\mu}_i$.) The resulting senescence
boundary condition, or ``birth" condition\footnote{A model where
  transition between some of the different classes occurs due to
  mutation would have an age boundary condition of the form
  $b_i(t,\bx,0) = \sum_k M_{ik} \int_0^\infty \hat{\beta}_k
  b_k(t,\bx,a) \ da$, where $M_{ik}$ is a matrix with mutation rates
  in the entries not on the main diagonal, and one minus the sum of
  those rates on the main diagonal.  The specifics of the entries of
  $M_{ik}$ would account for what mutations underlie the different
  phenotypes.  A version with linear progression through phenotype
  classes was used in \cite{MMStumor06}.}, is
\begin{equation} \label{bs_birth}
b_i(t,\bx,0) = \int_0^\infty \hat{\beta}_i(a,\ldots) b_i(t,\bx,a) \ da, \qquad \mbox{for } i=1,\ldots,N_b.
\end{equation}

We represent retention of inert cells using $N_b$ inert-cell classes governed by the conservation equations
\begin{equation}\label{n_i}
\frac{\partial n_i(t,\bx)}{\partial t} + \nabla \cdot {\bf J}_i^n =
\int_0^\infty \hat{\mu}_i(a,\ldots) b_i(t,\bx,a) \ da, \qquad
\mbox{for } i=1,\ldots,N_b.
\end{equation}

Conservation of substrate mass yields
\begin{equation}
\frac{\partial c_i}{\partial t} + \nabla \cdot {\bf J}_i^c = r_j, \qquad \mbox{for } j=1,\ldots,N_c,
\end{equation}
where $r_j$ denotes gain or loss of the $j$-th substrate concentration
through interactions with the biomass such as consumption or
excretion.  Assuming Fick's Law gives ${\bf J}_j^c = -D_j\nabla c_j$
for constants $D_j$.  The substrate masses are also subject to
advection, but the velocity is sufficiently slow that we can neglect
the advective contribution to the flux. Likewise, substrate material
diffuses several orders of
magnitude faster than the rates at which bacteria grow or advect,
allowing us to make a quasi-steady-state assumption so that
\begin{equation}
-D_j \nabla^2 c_j = r_j, \qquad \mbox{for } j=1,\ldots,N_c.
\end{equation}

We let $\vartheta_i(t,\bx,a)$ and $\rho_i(t,\bx,a)$ denote the volume
fraction per age and density per age relative to volume fraction, resp., of phenotypes
$i$, so that $b_i = \rho_i \theta_i$.  We assume incompressibility of
biomass with $\rho_i(t,\bx) \equiv \rho^*_i$ for positive constants
$\rho_i^*$.   We also assume inert cells have the same incompressibility
properties, and the same densities relative to volume fractions,
$\rho_i^*$, as active cells.  We let $\eta_i(t,\bx)$ denote the volume
fraction of inert phenotype $i$ cells, which is related to the
density of inert phenotype $i$ cells by $n_i = \rho_i^* \eta_i$.
We assume such cells all behave the same regardless of
phenotype, and track only the total
volume fraction of inert cells of all phenotypes, denoted by
$\mathcal{N}(t,\bx)$.  Equations (\ref{n_i}), rewritten as
\begin{equation}
\frac{\partial \eta_i(t,\bx)}{\partial t} + \frac{1}{\rho_i^*} \nabla
\cdot {\bf J}_i^n = \int_0^\infty \mu_i(a,\ldots) \vartheta_i(t,\bx,a)
\ da, \qquad \mbox{for } i=1,\ldots,N_b,
\end{equation}
become, after summing over $i$, the governing equation for $\mathcal{N}$,
\begin{equation}\label{N_t}
\frac{\partial \mathcal{N}(t,\bx)}{\partial t} + \sum_{i=1}^{N_b}
\frac{1}{\rho_i^*} \nabla \cdot {\bf J}_i^n = \mathcal{M}(t,\bx),
\end{equation}
where
\begin{eqnarray}
\mathcal{N}(t,\bx) &=& \sum_{i=0}^{N_b} \eta_i(t,\bx), \label{N} \\
\mathcal{M}(t,\bx) &=& \sum_{i=0}^{N_b} \int_0^\infty \mu_i(a,\ldots) \vartheta_i(t,\bx,a) \ da.
\end{eqnarray}
We require the biomass volume fractions to total to one so that
\begin{equation} \label{volumeConstraint}
\mathcal{N}(t,\bx) + \sum_{i=1}^{N_b} \int_0^\infty \vartheta_i(t,\bx,a) \ da = 1.
\end{equation}

Assuming that transport of biomass, including inert cells, is
governed by an advective process, with a volumetric flow $\bu(t,\bx)$
for all classes and ages, gives the fluxes ${\bf J}_i^b = \rho^*_i
\vartheta_i \bu$ for $i=1\ldots,N_b$.  Following
\cite{multispeciesBiofilm,DockeryKlapperSIAP01}, we assume that the
volumetric flow is stress driven according to
\begin{equation}
\bu = -\lambda \nabla p,
  \label{Darcy}
\end{equation}
where $p(t,\bx)$ is the pressure and $\lambda>0$ the Darcy
constant.  As in \cite{multispeciesBiofilm,DockeryKlapperSIAP01},
$p=0$ in $\Omega \backslash B_t$. Pressure is determined in order
to enforce incompressibility in response to growth (see below)
and hence~(\ref{Darcy}) can be viewed as a balance of growth-induced
stress against friction. Other choices of force balance are possible.

Substituting $b_i=\rho_i^* \vartheta_i$ and ${\bf J}_i^b = \rho^*_i
\vartheta_i \bu$ into equations (\ref{b_i}) gives, for
$i=1,\ldots,N_b$,
\begin{multline} \label{v_i}
\frac{\partial \vartheta_i}{\partial t} + \frac{\partial
  \vartheta_i}{\partial a} + \nabla \cdot (\bu \vartheta_i) =
-\mu_i(a,c_1,\ldots,c_{N_c},\vartheta_1(t,\bx,\cdot),\ldots,\vartheta_{N_b}(t,\bx,\cdot))
\vartheta_i(t,\bx,a) \\ +
f_i(\vartheta_1(t,\bx,a),\ldots,\vartheta_{N_b}(t,\bx,a)),
\end{multline}
where
\begin{multline}
\mu_i(a,c_1,\ldots,c_{N_c},\vartheta_1(t,\bx,\cdot),\ldots,\vartheta_{N_b}(t,\bx,\cdot))
= \\
\hat{\mu}_i(a,c_1,\ldots,c_{N_c},b_1(t,\bx,\cdot),\ldots,b_{N_b}(t,\bx,\cdot))
\end{multline}
and
\begin{equation}
f_i(\vartheta_1(t,\bx,a),\ldots,\vartheta_{N_b}(t,\bx,a)) =
\frac{1}{\rho_i^*}\hat{f}_i(b_1(t,\bx,a),\ldots,b_{N_b}(t,\bx,a)).
\end{equation}
The birth conditions (\ref{bs_birth}) become
\begin{equation}\label{vBirth}
\vartheta_i(t,\bx,0) = \int_0^\infty \beta_i(a,\ldots)
\vartheta_i(t,\bx,a) \ da, \qquad \mbox{for } i=1,\ldots,N_b.
\end{equation}
where
\begin{multline}
\beta_i(a,c_1,\ldots,c_{N_c},\vartheta_1(t,\bx,\cdot),\ldots,\vartheta_{N_b}(t,\bx,\cdot))
= \\
\hat{\beta}_i(a,c_1,\ldots,c_{N_c},b_1(t,\bx,\cdot),\ldots,b_{N_b}(t,\bx,\cdot)).
\end{multline}

Substituting ${\bf J}_i^n = \rho^*_i \eta_i \bu$ into equation
(\ref{N_t}) and using equation (\ref{N}) gives
\begin{equation} \label{N_t2}
\frac{\partial \mathcal{N}}{\partial t} = \mathcal{M} - \nabla \cdot \bu \mathcal{N},
\end{equation}
Integrating equation (\ref{v_i}) over age and summing over $i$ gives
\begin{multline} \label{bigCheese}
\underbrace{\frac{\partial}{\partial t} \left( \sum_{i=1}^{N_b}
  \int_0^\infty \vartheta_i(t,\bx,a) \ da \right)}_{=\nabla \cdot (\bu
  \mathcal{N})-\mathcal{M}} + \underbrace{\left( \sum_{i=1}^{N_b}
  \int_0^\infty \frac{\partial \vartheta_i(t,\bx,a)}{\partial a}  \ da
  \right)}_{= - \mathcal{B} \mbox{ as defined below} } = \\
- \underbrace{\nabla \cdot \left( \bu \sum_{i=1}^{N_b} \int_0^\infty
  \vartheta_i(t,\bx,a) \ da \right)}_{=\nabla \cdot
  (\bu(1-\mathcal{N})) } - \underbrace{\left( \sum_{i=1}^{N_b}
  \int_0^\infty \mu_i(a,\ldots) \vartheta_i(t,\bx,a) \ da
  \right)}_{=\mathcal{M}} \\
+ \underbrace{\left( \sum_{i=1}^{N_b} \int_0^\infty
  f_i(\vartheta_1(t,\bx,a),\ldots,\vartheta_{N_b}(t,\bx,a))\ da
  \right)}_{=\mathcal{F} \mbox{ as defined below}}.
\end{multline}

Using equations (\ref{volumeConstraint}) and (\ref{N_t2}),
we find that the first term in
the first line of equation (\ref{bigCheese}) is $\partial_t
(1-\mathcal{N}) = \nabla \cdot (\bu \mathcal{N}) -\mathcal{M}$.  For
the second term in the first line, we assume that $\vartheta_i$, for
$i=1,\ldots,N_b$, are sufficiently smooth, and that the corresponding
$\mu_i$ are bounded away from zero for $a$ large, so that each
$\vartheta_i$ will eventually decay exponentially to zero as $a
\rightarrow \infty$ (see section 7 in \cite{age-pwconst-paper}).  We
then obtain for the negative of the second term, using the age boundary
conditions defined by equation (\ref{vBirth}),
\begin{eqnarray}
\mathcal{B}(t,\bx) &=& \sum_{i=0}^{N_b} \vartheta_i(t,\bx,0) \nonumber \\
&=& \sum_{i=0}^{N_b} \int_0^\infty \beta_i(a,\ldots) \vartheta_i(t,\bx,a) \ da.
\end{eqnarray}
For the first term in the second line of equation (\ref{bigCheese}),
we again use equation (\ref{volumeConstraint}) to obtain $\nabla \cdot
(\bu(1-\mathcal{N}))$.
Recall that the second to last term of equation (\ref{bigCheese}) is
just $\mathcal{M}(t,\bx)$ and set the last term to
\begin{equation}
\mathcal{F}(t,\bx) = \sum_{i=0}^{N_b} \int_0^\infty f_i \ da.
\end{equation}
We note that $\mathcal{F}$ is not generally identically zero; a
similar sum over all phenotypes of the integrals over all ages of the
net changes between phenotypes, $\hat{f}$, is conserved to be zero.
However, because the densities relative to volume fractions,
$\rho_i^*$, are not identical, we generally only have $\mathcal{F}
\equiv 0$ when $\rho_i^* = \rho^*$ for some constant $\rho^*$ and for
all $i=1,\ldots,N_b$.  We rewrite equation (\ref{bigCheese}) more
compactly as
\begin{equation} \label{littleCheese}
\nabla \cdot \bu = \mathcal{B}(t,\bx) + \mathcal{F}(t,\bx),
\end{equation}
the incompressibility relation for our system.

Substituting $\bu = -\lambda \nabla p$ results in an equation for the
pressure, namely
\begin{equation} \label{pressure}
-\lambda \nabla^2 p = \mathcal{B}(t,\bx) + \mathcal{F}(t,\bx), \qquad \mbox{in } B_t.
\end{equation}
Distributing the divergence operator, and again using $\bu = -\lambda
\nabla p$ along with equation (\ref{littleCheese}), gives us $\nabla
\cdot (\bu \vartheta_i) = -\lambda \nabla p \cdot \nabla \vartheta_i +
\vartheta_i (\mathcal{B} + \mathcal{F})$, so that equation (\ref{v_i})
can be rewritten, for $i=1,\ldots,N_b$,
\begin{equation}
\frac{\partial \vartheta_i}{\partial t} + \frac{\partial
  \vartheta_i}{\partial a} - \lambda \nabla p \cdot \nabla \vartheta_i
= -\mu_i \vartheta_i + f_i - \vartheta_i (\mathcal{B} + \mathcal{F}),
\end{equation}
Similarly, we rewrite equation (\ref{N_t2}) as
\begin{equation} \label{N_t3}
\frac{\partial \mathcal{N}}{\partial t} = \mathcal{M} + \lambda \nabla
p \cdot \nabla \mathcal{N} - \mathcal{N} (\mathcal{B} + \mathcal{F}).
\end{equation}
We see from equation (\ref{pressure}) that $p$ is proportional to
$\lambda^{-1}$, so that $\lambda \nabla p$ is independent of
$\lambda$.  Consequently, $\vartheta_i$ and $\mathcal{N}$ are
independent of $\lambda$, allowing us to set $\lambda=1$.

We impose periodic and other boundary conditions similar to what was
done in \cite{multispeciesBiofilm} to obtain the complete model, for
$i=1,\ldots,N_b$ and $j=1,\ldots,N_c$,
\begin{subequations} \label{system}
\begin{equation}
\frac{\partial \vartheta_i}{\partial t} + \frac{\partial
  \vartheta_i}{\partial a} - \nabla p \cdot \nabla \vartheta_i =
-\mu_i \vartheta_i + f_i - \vartheta_i (\mathcal{B} + \mathcal{F}),
\qquad \bx \in B_t, t>0, a>0,
\end{equation}
\begin{align}
\vartheta_i(t,\bx,0) = \int_0^\infty \beta_i(a,\ldots)
\vartheta_i(t,\bx,a) \ da,  \qquad & \bx \in B_t, t>0, \\
\frac{\partial \vartheta_i}{\partial z} = 0, \qquad & \bx \in \Gamma_B, t\geq 0, a >0, \\
\vartheta_i(0,\bx,a) = \vartheta_i^0(\bx,a), \qquad& \bx \in B_t, a\geq 0,
\end{align}
\begin{align}
\frac{\partial \mathcal{N}}{\partial t} - \nabla p \cdot \nabla
\mathcal{N} = \mathcal{M}  - \mathcal{N} (\mathcal{B} + \mathcal{F}),
\qquad& \bx \in B_t, t>0, \\
\frac{\partial \mathcal{N}}{\partial z} = 0,  \qquad & \bx \in \Gamma_B, t\geq 0, \\
\mathcal{N}(0,\bx) = \mathcal{N}^0(\bx), \qquad & \bx \in B_t,
\end{align}
\begin{align}
-\nabla^2 p  = \mathcal{B} + \mathcal{F}, \qquad& \bx \in B_t, t\geq 0, \\
p = 0, \qquad& \bx \in \Gamma_t, t\geq 0, \\
\frac{\partial p}{\partial z} = 0, \qquad& \bx \in \Gamma_B, t\geq 0,
\end{align}
\begin{align}
-D_j \nabla^2 c_j = r_j, \qquad& \bx \in \Omega, t>0, \\
 r_j = 0, \qquad& \bx \in \Omega\backslash B_t,\\
c_j =  c_j^*, \qquad& \bx \in \Gamma_{H_b}, t\geq 0, \\
\frac{\partial c_j}{\partial z} = 0, \qquad& \bx \in \Gamma_B, t\geq 0.
\end{align}
The normal velocity of the interface $\Gamma_B$ is given by
\begin{equation}\label{fullDnormal}
-\nabla p \cdot {\bf n} = -\frac{\partial p}{\partial n},
\end{equation}
where ${\bf n}$ is the unit outward normal of $\Gamma_B$.
\end{subequations}

We make particular choices of functions $\beta$ and $\mu$ as follows.
To reflect the diminished new-cell production by senescent cells
discussed in \cite{StewartEtAl2005}, we define senescence as a
function of age, $\sigma(a)$, such that $\sigma(0)=0$ and $\sigma(a) \rightarrow 1$
as $a \rightarrow \infty$, and incorporate $\sigma(a)$ into $\beta(a,c)$
and $\mu(a,c)$,
\begin{subequations}\label{betaMu}
\begin{align}
\beta(a,c) &= \beta_0(c) \left(1-\sigma(a)\right),  \\
\mu(a,c)&=\mu_0(c) \sigma(a).
\end{align}
\end{subequations}
We neglect the $c$ dependence of $\mu_0$, and choose
\begin{subequations}\label{sigmaBeta}
\begin{align}
\sigma(a) &= \frac{a}{a^*+a}, \qquad a\geq 0, \label{sa}\\
\beta_0(c) &= \frac{\psi \, c}{k+c}, \label{b0}
\end{align}
\end{subequations}
where $a^*$ is the senescence age scale, and, following
\cite{multispeciesBiofilm}, $\psi$ is the maximum growth
rate (with units of inverse age) and
$k$ is the Monod saturation constant (with units of concentration).
Oxygen uptake has the form
\begin{equation}
r(c,\vartheta(t,\bx,\cdot)) = \frac{\xi \, c}{k+c} \int_0^\infty (1-\sigma(a)) \vartheta(t,\bx,a) \ da,
\end{equation}
where $\xi$ is the maximum uptake rate. 

\section{Non-dimensionalization}
\label{nondim}

We simplify in the following to $N_b=N_c=1$, i.e., restrict to one
active phenotype and one substrate, and drop indexing subscripts.
Note now that $\mathcal{N}=\eta(t,\bx)$ and
$\mathcal{M}=\int_{0}^{\infty}\mu(a,\ldots)\vartheta(t,\bx,a)\,da$.
Also note that $f=\mathcal{F}=0$. We will continue to assume
that $\beta=\beta(a,c)$ and $\mu=\mu(a,c)$.

Let $\bar{\beta}$ be a typical value of $\beta(a,c)$ and
let $\bar{\mu}$ be a typical value of $\mu(a,c)$.
Choosing a characteristic time scale $T=1/\bar{\beta}$,
age scale $A=1/\bar{\mu}$, and temporarily reintroducing the
friction coefficient $\lambda$, we nondimensionalize
according to $\tilde{t}=t/T$, $\tilde{a}=a/A$, $\tilde{x}=x/L$,
and $\tilde{\beta}=\beta T$, $\tilde{\mu}=\mu A$, $\tilde{c}=c/c^*$,
$\tilde{r}=r/r(c^*)$, $\tilde{p}=p(\lambda T/L^2)$,
$\tilde{\vartheta}=\vartheta A$, $\tilde{\eta}=\eta$.
Here $L$ is a (problem-dependent) characteristic system
length scale.

Substituting into the system (\ref{system}) and dropping tildes, we obtain
\begin{subequations} \label{system2}
\begin{equation}
\frac{\partial \vartheta}{\partial t} + \Lambda \frac{\partial\vartheta}{\partial a}
   - \nabla p \cdot \nabla \vartheta =
-\Lambda \mu \vartheta - \vartheta\int_{0}^{\infty}\beta(a,c)\vartheta(a)\,da,
\qquad \bx \in B_t, t>0, a>0,
   \label{theta_eqn}
\end{equation}
\begin{align}
\vartheta(t,\bx,0) = \int_0^\infty
      \beta(a,c) \vartheta(t,\bx,a) \ da,  \qquad & \bx \in B_t, t>0, \label{theta_0} \\
\frac{\partial \vartheta}{\partial z} = 0, \qquad & \bx \in \Gamma_B, t\geq 0, a >0, \\
\vartheta(0,\bx,a) = \vartheta_0(\bx,a), \qquad& \bx \in B_t, a\geq 0,
\end{align}
\begin{equation}
\frac{\partial\eta}{\partial t} - \nabla p \cdot \nabla\eta
   = \Lambda \int_{0}^{\infty}\mu(a,c)\vartheta(a)\,da -
     \eta\int_{0}^{\infty}\beta(a,c)\vartheta(a)\,da, \quad \bx \in B_t, t>0, \label{eta_eqn}
\end{equation}
\begin{align}
\frac{\partial\eta}{\partial z} = 0,  \qquad & \bx \in \Gamma_B, t\geq 0, \\
\eta(0,\bx) = \eta_0(\bx), \qquad & \bx \in B_t,
\end{align}
\begin{align}
\nabla^2 p  = -\int_{0}^{\infty}\beta(a,c)\vartheta(a)\,da, \qquad& \bx \in B_t, t\geq 0, \\
p = 0, \qquad& \bx \in \Gamma_t, t\geq 0, \\
\frac{\partial p}{\partial z} = 0, \qquad& \bx \in \Gamma_B, t\geq 0,
\end{align}
\begin{align}
\nabla^2 c = -Gr, \qquad& \bx \in \Omega, t>0, \label{c_eqn} \\
 r = 0, \qquad& \bx \in \Omega\backslash B_t,\\
c =  1, \qquad& \bx \in \Gamma_{H_b}, t\geq 0, \\
\frac{\partial c}{\partial z} = 0, \qquad& \bx \in \Gamma_B, t\geq 0.
\end{align}
\end{subequations}
Here $G=L^2 r(c^*)/(c^*D)$ and thus $1/\sqrt{G}$, the active
layer depth, is a non-dimensional measure of the depth (scaled by system size)
to which substrate can penetrate into the biofilm before
it is consumed. Likewise
\begin{equation}
\Lambda=\frac{\bar{\mu}}{\bar{\beta}}
\end{equation}
 is a non-dimensional ratio of characteristic deactivity time to characteristic
reproduction time.

The non-dimensional forms of equations (\ref{sa}) and (\ref{b0}) are
\begin{subequations}
\begin{align}
\sigma(a) &= \frac{a}{S+a}, \qquad a\geq 0,\\
\beta_0(c) &= \frac{P \, c}{K+c},
\end{align}
\end{subequations}
where $S=a^*/A$ is a comparison of senescence age with system
age scale, $K=k/c^*$ is a measure of saturation level
(large $K$ means substrate-limited behavior and small $K$
indicates growth-limited behavior), and $P=\psi/\bar{\beta}$
is a measure of maximum to typical yield.

The magnitude of $\Lambda$ may depend on location within the biofilm.
We identify two regimes. First, near the top of the biofilm,
in particular within the active layer, $c$ is ${\mathcal O}(1)$ and we
can then generally expect for a viable biofilm
that $\bar{\beta}$ be large compared
to $\bar{\mu}$, i.e., $\Lambda$ small. In this case advective terms
dominate in equations~(\ref{theta_eqn}) and~(\ref{eta_eqn}) over
the death terms. If advection is unimportant, i.e.,
$\nabla p\cdot\nabla\vartheta$ is small, then age scale is
determined by the second term of~(\ref{theta_eqn}) which
then requires $\partial/\partial a\sim\Lambda^{-1}$. Such scaling
is in fact observed within the biofilm active layer, see
Section~\ref{computations}.

Second, beneath the active layer, equation (\ref{c_eqn})
indicates exponential decay (in space) of $c$. Hence, below
a sharp transition region from the active layer, we can expect
$\bar{\beta}$ to be small compared to $\bar{\mu}$, i.e., $\Lambda$ large.
In this case at first glance equation (\ref{theta_eqn})
indicates $\vartheta$ has an $\mu$-governed decaying age structure.
There is a subtlety here however. Large $\Lambda$ in~(\ref{theta_eqn})
suggests an approximately exponential age profile of the
form $\vartheta(a)=\vartheta(0)\exp(-\mu a)$. But such a form
is inconsistent with~(\ref{theta_0}), which does not allow
dependence on $\mu$, unless $\vartheta(0)=0$. In other words,
for large $\Lambda$ the birth term is insufficient to introduce
enough new cells to overcome death and so an active population
is not viable. Having said this, however, we will observe a
$\mu$-determined exponential age structure develop in the
deeper parts of the biofilm, see Section~\ref{computations}.
The reason for this is that in the lower layer of the biofilm, where
the population is barely viable, the birth rate has decreased to the point
that it is only just balancing death. Hence condition (\ref{theta_0}),
which requires that an exponential age structure be controlled
by $\beta$, also implies that exponential age structure be
determined by $\mu$.

\section{Spatially Homogeneous Steady-State Age Distributions}
\label{steady-state}

We assume spatial homogeneity and temporal stationarity, i.e.,
$\vartheta=\vartheta(a)$, $\eta=\eta(a)$, on $-\infty<z<\infty$.
(We note that pressure gradients
within the biofilm and hence advection are generally weak
within inactive regions.) Then $\vartheta$, $\eta$ satisfy
\begin{eqnarray}
\frac{d\vartheta}{da} & = & -(\mu+\vartheta_0)\vartheta, \label{th}\\
\vartheta_0 & = & \int_0^{\infty}\beta\vartheta\,da, \label{th0} \\
\eta & = & 1 -\int_{0}^{\infty}\vartheta(a)\,da, \nonumber
\end{eqnarray}
where $\vartheta_0=\vartheta(0)$.
This system is unphysical in that it requires unbounded velocities
(the pressure gradient takes the form $p_z=C_1z+C_2$) to enforce
incompressibility, but it is nevertheless
useful for illustrative  purposes.

The solution to equation (\ref{th}) is
\begin{equation}
\vartheta(a) = \vartheta_0e^{-\int_0^a\mu\,da'}e^{-\vartheta_0a}.
  \label{theta}
\end{equation}
Thus equation (\ref{th0}) implies the condition
\begin{equation}
\vartheta_0 = \int_0^{\infty}\beta\vartheta_0
      e^{-\int_0^a\mu\,da'}e^{-\vartheta_0a}\,da.
\end{equation}
In order to have a nontrivial solution, we require $\vartheta_0$ to satisfy
\begin{equation}
1 =  \int_0^{\infty}\beta e^{-\int_0^a\mu\,da'}e^{-\vartheta_0a}\,da,
    \label{condition}
\end{equation}
if possible. If this is not possible, then $\vartheta_0=\vartheta(a)=0$ is the
only solution.

The choice of $\mu$ and $\beta$ independent of $a$ allows
a particular transparence. In this case condition (\ref{condition}) becomes
\begin{equation}
1 =  \int_0^{\infty}\beta e^{-(\mu+\vartheta_0)a}\,da,
\end{equation}
which has a solution with $\vartheta_0>0$ if
\begin{equation}
\int_0^{\infty}\beta e^{-\mu a}\,da>1,
\end{equation}
i.e., if new cells can be produced sufficiently fast to
replace aging (and dying) ones. Note that this condition
cannot be satisfied for $\mu$ sufficiently large or
$\beta$ sufficiently small (in which
case $\vartheta_0=0$ necessarily).
Now if we write $\vartheta_0=\hat{\vartheta}_0-\mu$, then
$\hat{\vartheta}_0$ solves
\begin{equation}
1 = \int_0^{\infty}\beta e^{-\hat{\vartheta}_0a}\,da,
\end{equation}
that is, $\hat{\vartheta}_0=\beta$. Equation (\ref{theta}) becomes
\begin{equation}
\vartheta(a) = (\beta-\mu)e^{-\beta a}.
\end{equation}
Fecundity $\beta$ fixes the profile of the age structure
though age distribution amplitude depends on both $\beta$ and $\mu$.
Large $\beta$ results in a
steep age profile with amplitude almost independent of $\mu$.
Small $\beta$ results in a flat age profile. However
we note that the viability boundary (in parameter space) occurs
at $\beta=\mu$. For marginally viable populations $\beta=\mu+\epsilon$
and hence the population age profile is exponential with decay rate
approximately $\mu$. Note that $\mu$ gives the slowest possible
rate of decay. With regards to a biofilm model, if we assume that
$\beta$ decreases with decreasing $c$, then 
age structure should flatten deeper down into the biofilm.
In fact, we expect an abrupt transition from steep to flat profile
as we pass through the active layer.

\begin{figure}[t]
\begin{center}
\subfigure[Response of $\ln(\theta(a))$ to changes in $\bar{\beta}$. ]{
\includegraphics[height=2.2in]{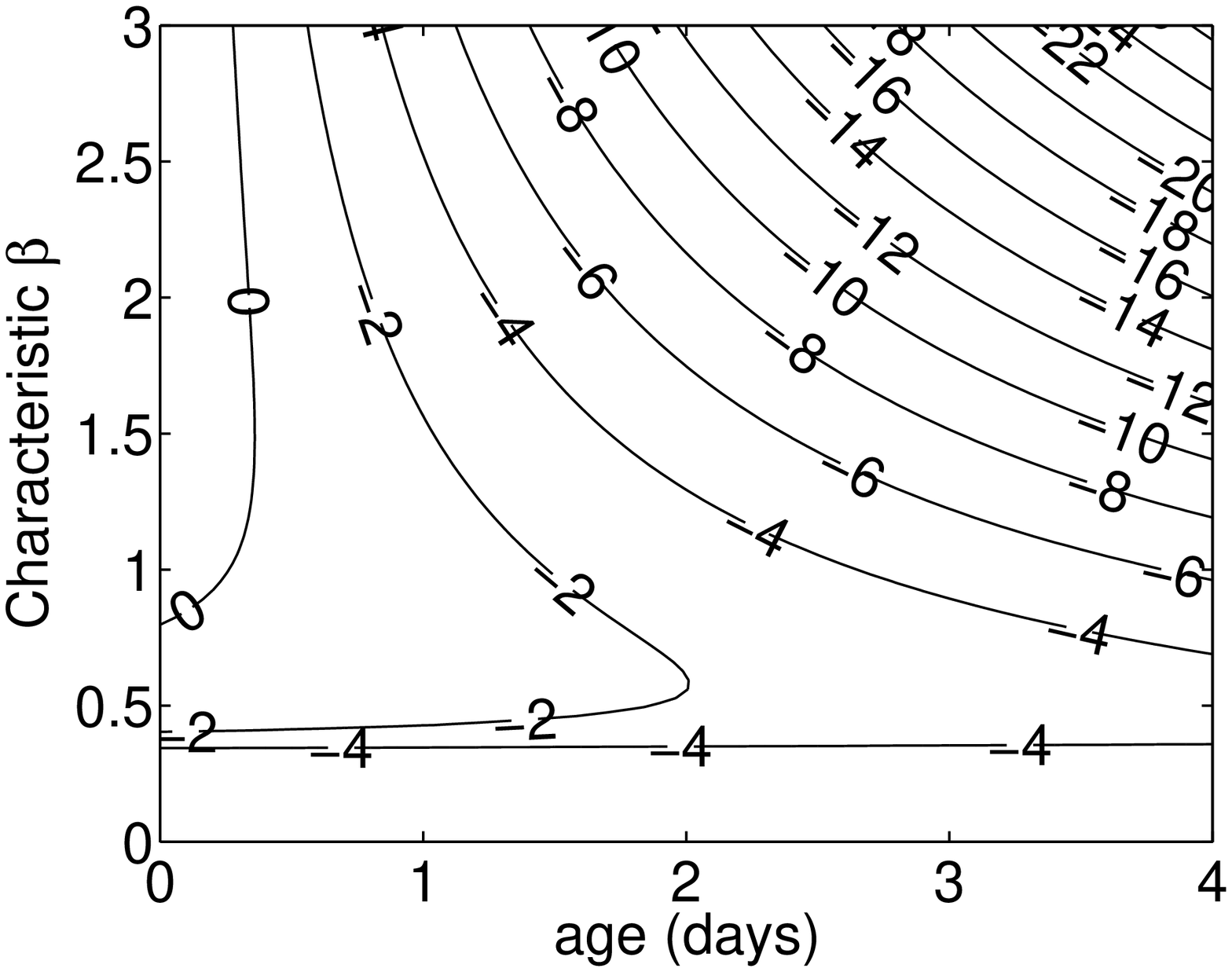}
}
\subfigure[Response of $\eta$ to changes in $\bar{\beta}$.]{
\includegraphics[height=2.2in]{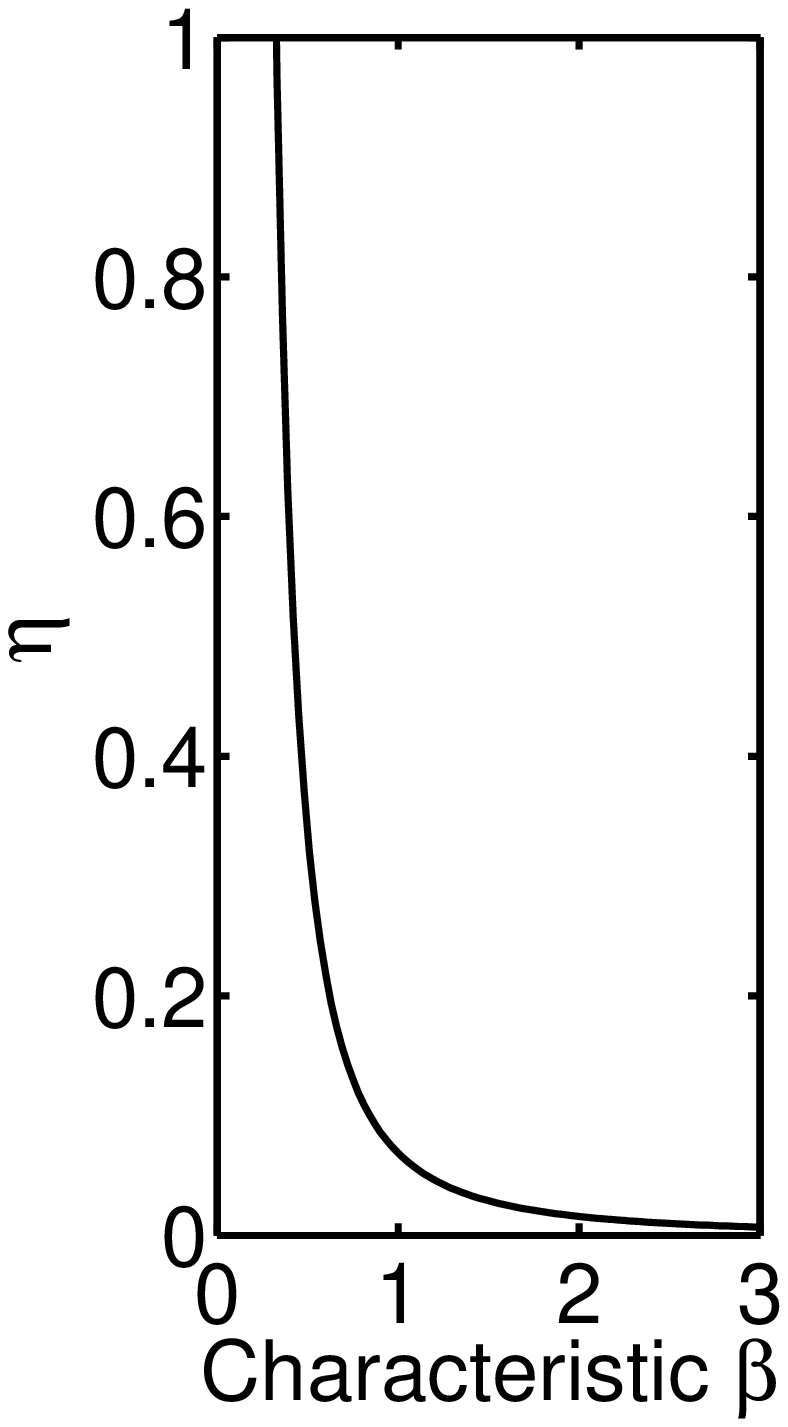}
}
\caption{Response of steady-state solutions to changes in the characteristic reproduction rate,
  $\bar{\beta} = \int_0^{a^*} \beta(a,c) \, da$.}
\label{nonDimFig}
\end{center}
\end{figure}

Returning to our specified forms of $\beta(a,c)$ and $\mu(a,c)$ we have
\begin{subequations}
\begin{align}
\vartheta(a) &= \vartheta(0)\left(\frac{S}{a+S} \right)^{-\mu_0(1)S}
                             e^{-\mu_0(1)a}e^{-\vartheta(0)a/\Lambda}, \label{SHSS} \\
\eta &= \Lambda \frac{\mu_0(1)}{\vartheta(0)}
             \int_0^{\infty}\frac{a}{a+S}\vartheta(a)\,da.
\end{align}
\end{subequations}
Going back for a moment to dimensional variables, we use days as units of time, take $\mu_0 =
0.25$, and vary $\beta_0(c)$ to induce changes in a characteristic
reproduction rate, $\bar{\beta} = \int_0^{a^*} \beta(a,c) \, da$.  We
set $a^*=0.5$, which accounts for a loss of 1\% vitality per division
(occurs on average every 0.02 days for {\em E. coli}).  Figure
\ref{nonDimFig} shows the response of the steady-state solutions to
changes in  $\bar{\beta}$.   We note that the steady-state active-cell
distribution is zero when $\bar{\beta}$ is less than roughly 0.33.

\section{Computational Results}
\label{computations}

In this section we present computational results for one spatial dimension
(height of the biofilm), and explicit age structure representing cell
senescence, for the dimensional system (\ref{system}).  The height of
the biofilm, $\Gamma_t$, is regulated using an
erosion term at the biofilm/substrate interface (a standard devise
in biofilm models, see e.g. \cite{GujerWanner90}) given by
\begin{equation} \label{Gamma_t}
\frac{\partial \Gamma_t}{\partial t} = -\frac{\partial p}{\partial z} \bigg\vert_{z=\Gamma_t} - \alpha \Gamma_t^2,
\end{equation}
where $\alpha$ is the erosion coefficient.

\begin{figure}[t]
\begin{center}
\hspace{-0.24in}
\includegraphics[height=1.6in]{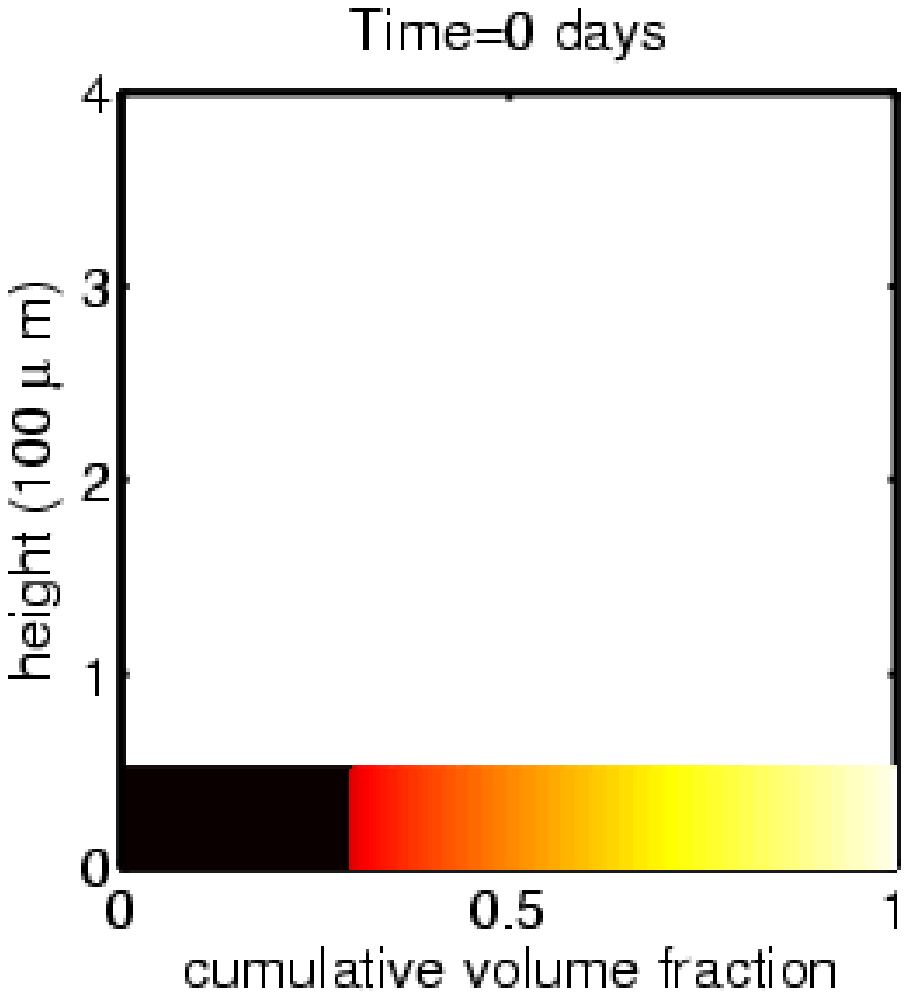}
\includegraphics[height=1.6in]{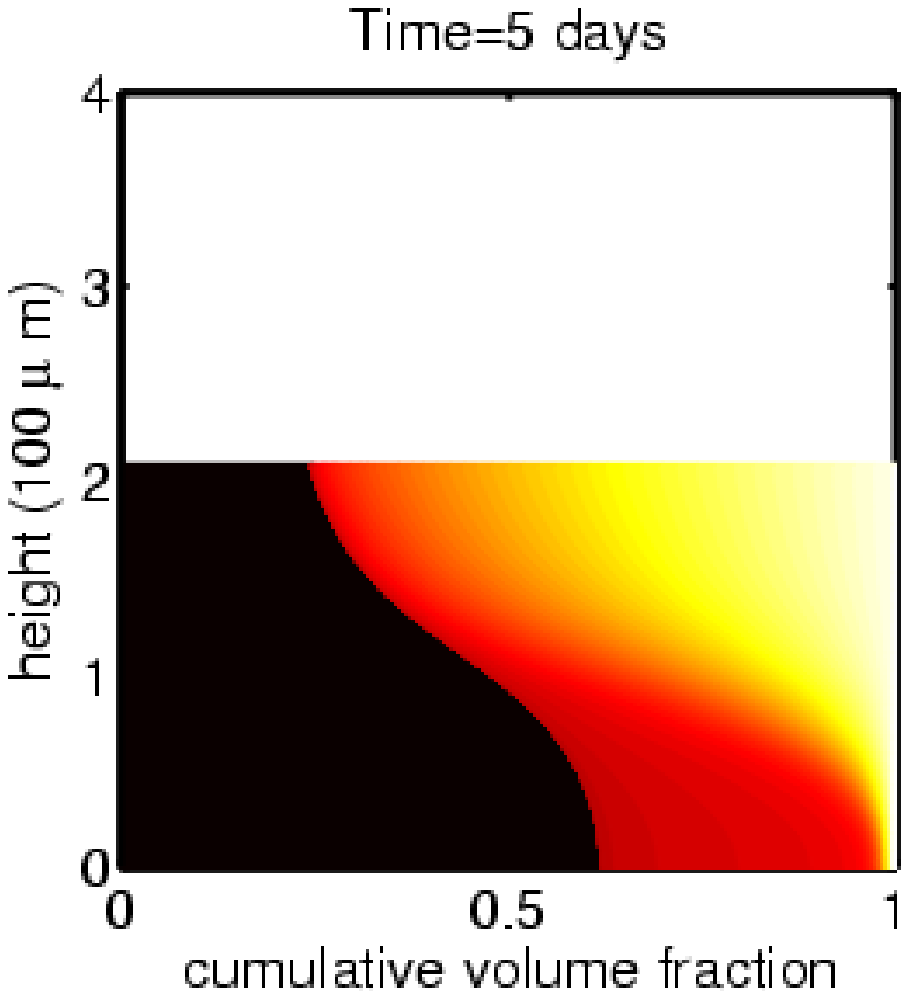}
\includegraphics[height=1.6in]{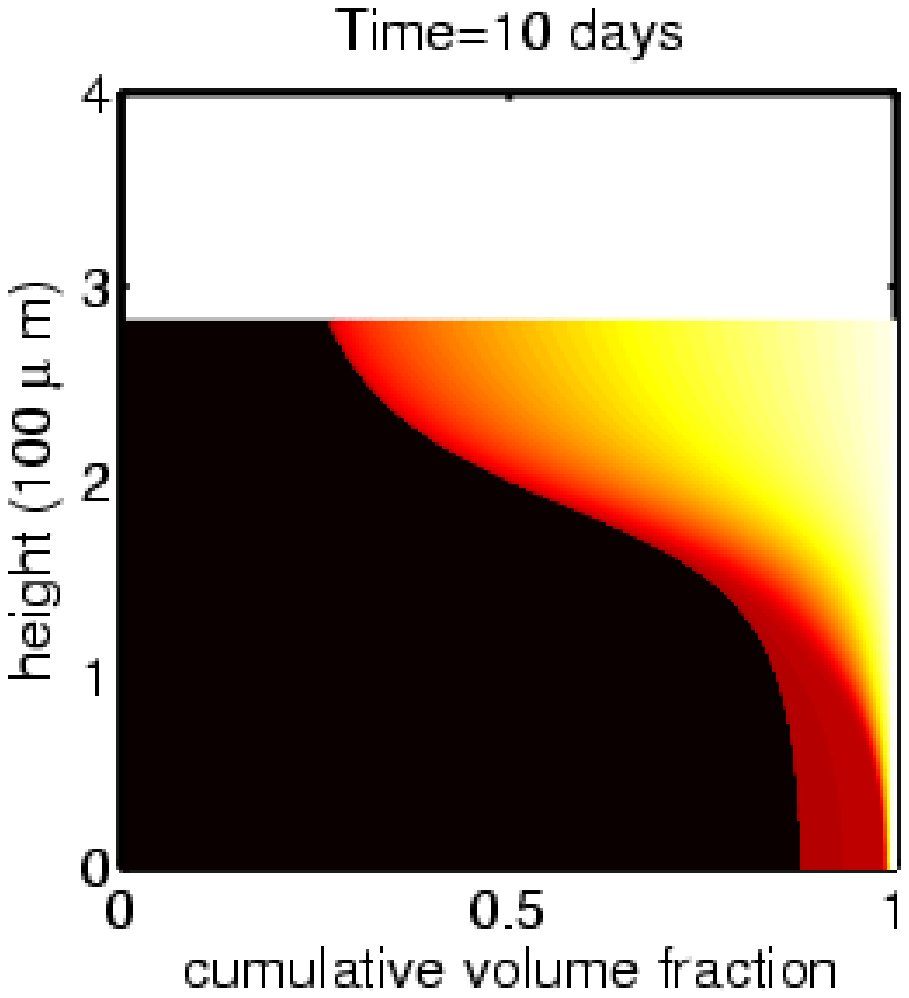} \\
\includegraphics[height=1.6in]{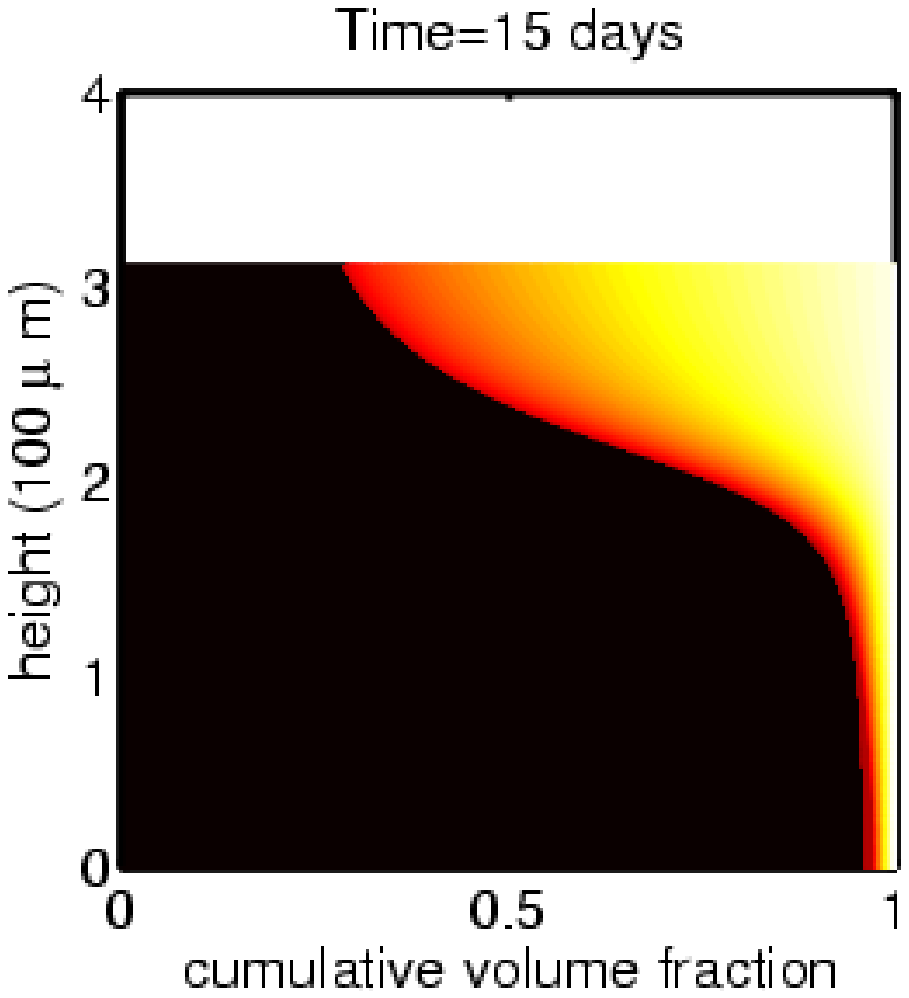}
\includegraphics[height=1.6in]{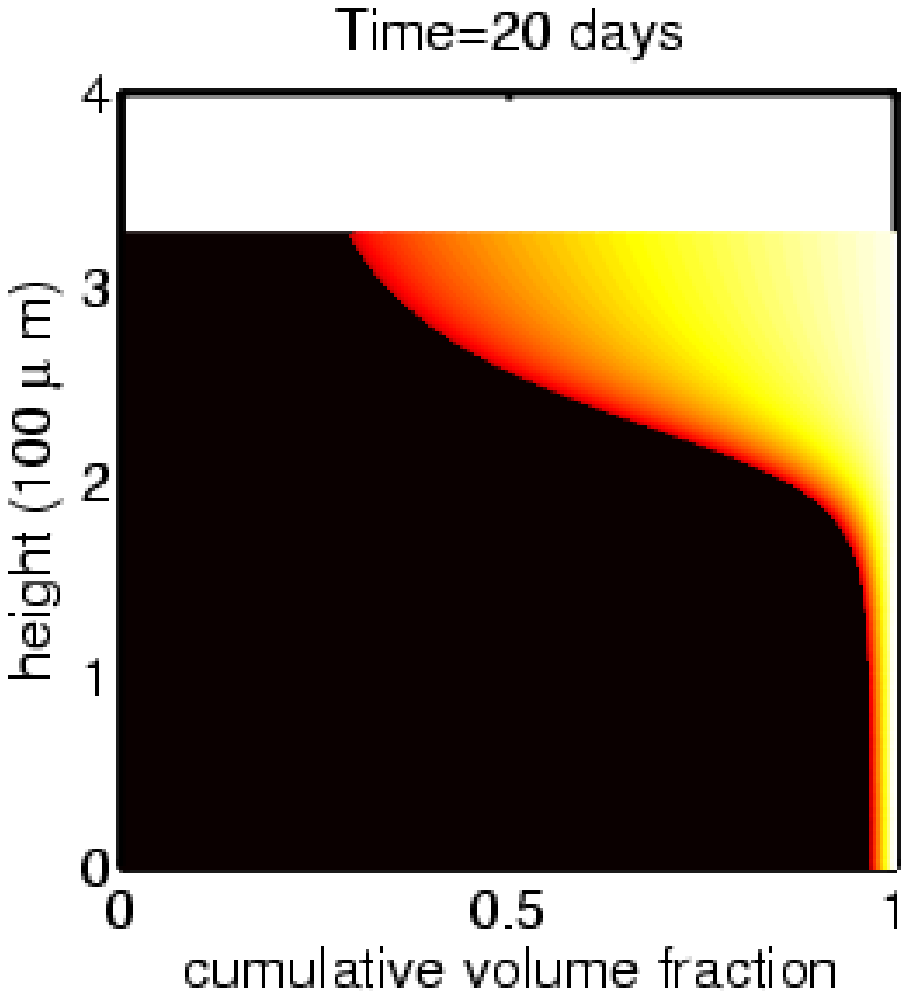}
\includegraphics[height=1.6in]{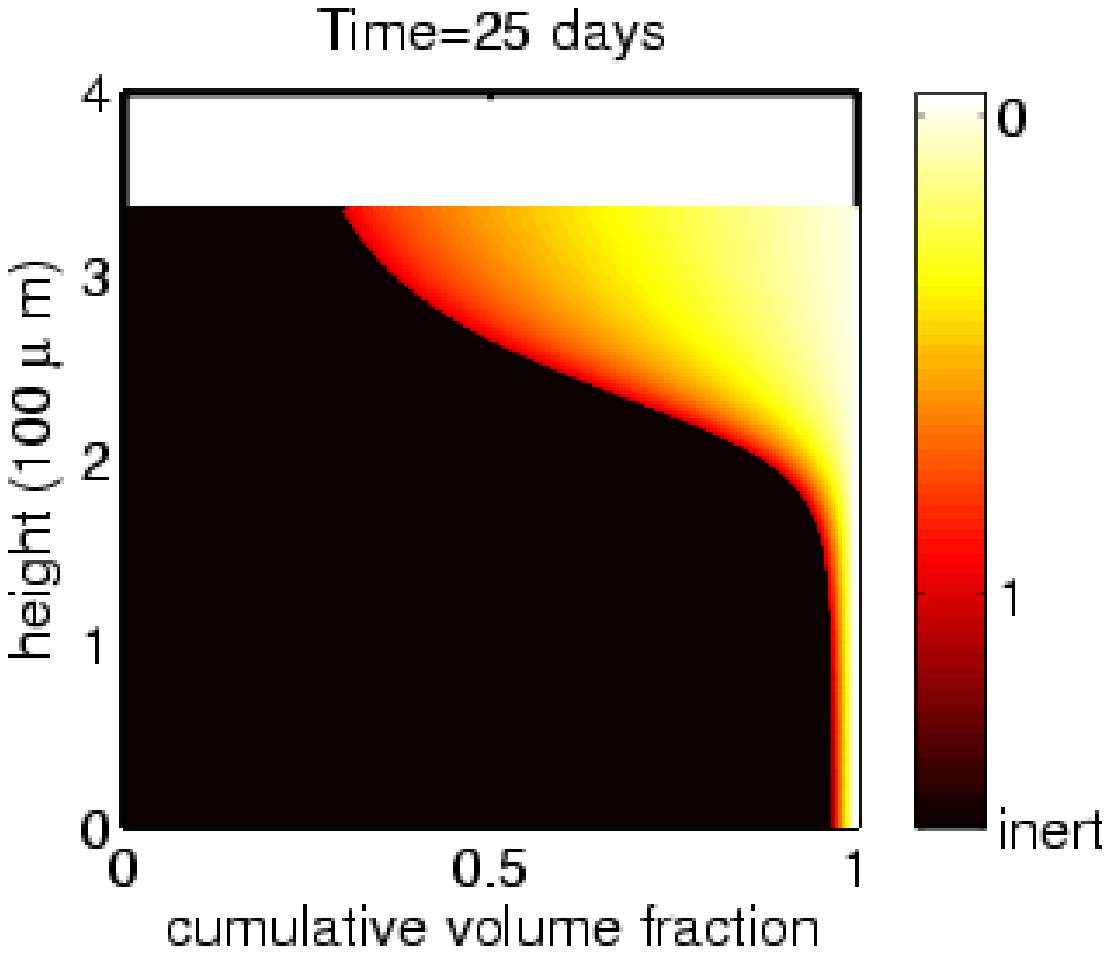}
\caption{Biofilm dynamics from initial colonization to steady state. The height of the colored area represents
the height of the biofilm, including both active and inert bacteria.
Color represents cell state: black represents inert cells, and a spectrum from
off-white to yellow to orange to red represents senescence of a cell
of a given age, $\sigma(a)$.  The horizontal width of a color
constitutes the volume fraction of cells of the corresponding senescence.}
\label{biofilmStart}
\end{center}
\end{figure}

For the computations presented in this section, we consider again the case
of $N_b=N_c=1$.  We take as the initial
condition a biofilm with a height of $\Gamma_t(0) = 50 \mu\mbox{m}$ and with an age
distribution that is initially the same for all heights,
$\vartheta(0,z,a) = 0.35*\max ( 1-\frac{a}{4},0 )$ for
$0\leq z \leq 50 \mu\mbox{m}$.  This piecewise linear function, when converted
from age structure to senescence structure (recall $\sigma(a) =
\frac{a}{a^*+a}$) \footnote{Computations with different $\sigma(a)$,
  namely $\sigma(a) = 1-\exp(-a/a^*)$, $\sigma(a) =
  \max(\frac{a}{4a^*}, 0.999)$ and $\sigma(a) = \max(\frac{a}{a^*},
  0.999)$, and with
  the same parameters as
  in this section, except for $\psi$, give qualitatively similar
  results.  We need to change $\psi$ since the different areas under
  the curves of $\sigma(a)$ give substantially different total new
  cell production.}, closely approximates the senescence structure at
the top of the biofilm when it is near steady state and thus
represents a situation where a new area is being colonized by material
from the top of a mature biofilm when it is near steady state. The
motivation is that a young biofilm may be formed by colonization of
cells detached from the upper region of an upstream, mature biofilm.

We use a time unit equal to one day.  As a result, we take the
senescence time scale to be $a^*=0.5$, as was done in Section \ref{steady-state}.  We use the division time for
{\em E. coli}, which is roughly every 30 minutes.

We set the erosion parameter to
be $\alpha=0.03$, the distance between $\Gamma_t$ and $\Gamma_{H_b}$
to be $H_b=37.5 \mu\mbox{m}$, and the parameters for the various
functional forms to be $\mu_0 = 0.25$, $k=0.05$, $\psi=2$, $\xi = 3$,
and $c^*=1$.  These parameter values are of the same order of
magnitude of those used in \cite{multispeciesBiofilm}, with
modifications due to the inclusion of age structure in the model
equations.

Results of the computations with the above parameters are displayed in
Figure \ref{biofilmStart}.  The height of the colored area indicates
the height of the biofilm, including both active and inert bacteria.
Color represents cell state: black designates inert cells, and a spectrum from
off-white to yellow to orange to red designates senescence of a cell
of a given age, $\sigma(a)$.  The horizontal width occupied by a color
indicates the volume fraction of cells of the corresponding senescence.

\begin{figure}
\begin{center}
\includegraphics[height=3in]{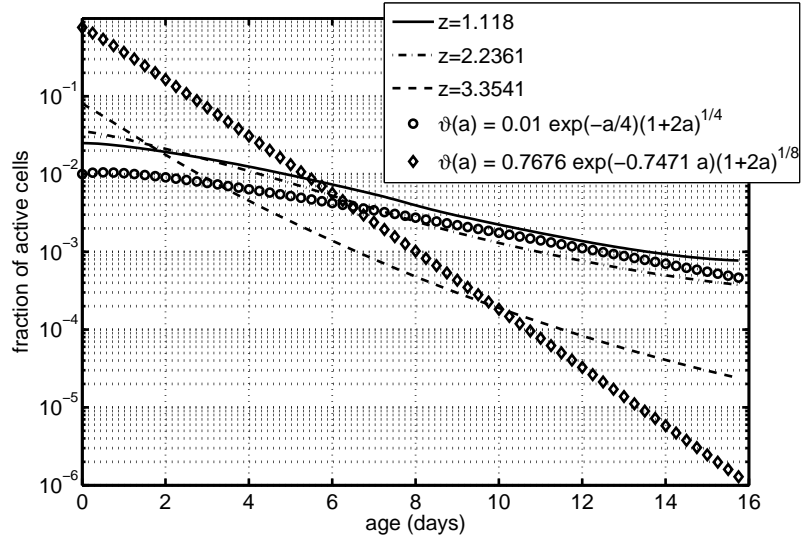}
\caption{Normalized age
distributions, ignoring the inert cell populations, one third of the way from the bottom, two
thirds of the way from the bottom, and at the top of the biofilm at
time $t=25$ days.    The plot
of $\vartheta(a)=0.01 \, \exp(-a/4)(1+2a)^{1/4}$ is the large $\Lambda$ limit of equation (\ref{SHSS}).   The coefficient of 0.01 governing the magnitude of the curve is chosen for ease of comparison.   The plot of  $\vartheta(a)=0.7676 \, \exp(-0.7471\, a)(1+2a)^{1/8}$ is the re-normalized steady-state, given an oxygen concentration of $c=0.5581$, in the absence of  advection.   Differences between this curve and the computed solution at the top of the biofilm illustrate the role of advection, including the upward flow of a relatively greater proportion of inert and senescent cells}
\label{ageDists}
\end{center}
\end{figure}

The biofilm tends to a steady state, as discussed in
Section \ref{steady-state}, consisting of an active layer at the top
and passive layer appearing as a stalk.  It is already understood that
this physical structure provides a form of protection for the bacteria
population as a whole \cite{ChamblessHuntStewart06}.  The question remains: how does senescence and
the corresponding resistance to antimicrobial challenge fit into the
overall defensive strategy of bacteria?

To obtain an answer, we first consider the normalized age
distributions, ignoring the inert cell populations, one third of the way from the bottom, two
thirds of the way from the bottom, and at the top of the biofilm at
time $t=25$ days.  These distributions are shown in Figure
\ref{ageDists}.

As we descend the biofilm down through the active
layer and into the passive layer, we expect $\Lambda$ to increase so
that equation (\ref{theta_eqn}) approaches, in steady state, equation \ref{SHSS}.   The plot
of $\vartheta(a)=0.01 \, \exp(-a/4)(1+2a)^{1/4}$ highlights the convergence toward the shape of the curve of the large $\Lambda$ limit of equation (\ref{SHSS}).   The coefficient of 0.01 governing the magnitude of the curve is chosen for ease of comparison.  At the top of the biofilm at steady-state, the oxygen concentration is approximately $c=0.5581$.   Using this value, our specific functional forms defined in equations (\ref{betaMu})-(\ref{sigmaBeta}), and equations (\ref{theta}) and (\ref{condition}), we obtain a value of $\theta_0 =  0.6221$.   We re-normalize the age distribution to total one so that equation (\ref{theta}) has the specific form  $\vartheta(a) \approx 0. 7676\, \exp(-0.7471 \, a)(1+2a)^{1/8}$.  This represents the situation when there is no advection.   Differences between this function and the graph of the computed steady-state age distribution at the top of the biofilm reflect the role of advection, including the upward flow of material with a relatively higher proportion of inert and senescent cells.

We find the expected result, as discussed in Sections
\ref{nondim} and \ref{steady-state}, that the age distributions broaden as we
go from the active to the passive layers within the biofilm.
In the inactive region, the profile matches that of an approximately
growth-death balanced population. This is to be expected in
an erosion maintained steady state -- some growth must occur all the
way to the bottom of the biofilm. A non-viable zone  does not form
in the presence of erosion because such a zone does not result in
any growth induced pressure and hence does not increase expansive
velocity. We remark that it is possible that $\mu$-dominated age
structure in the inactive region of the biofilm is thus a byproduct
of erosion. A more general (and realistic) biofilm model would
allow for the possibility of mechanical detachment; this
would require a much more elaborate set-up than the one used here
(i.e., mechanical stress coupling in three dimensions). It is still plausible that a non-viable zone would lead to detachment and so
we posit that age structure would not change.

Finally, we extend the computation to include the effects of
antimicrobial challenge.  We assume the antimicrobial agent has a source at the
bulk-substrate interface $\Gamma_{H_b}$, that it diffuses on a fast time
scale compared to growth, and, for simplicity, 
that it is not degraded by the biofilm.  Consequently,
the antimicrobial saturates the biofilm essentially
instantaneously and thus we
can model the effects of antimicrobial challenge by modifying the death modulus $\mu$
rather than adding an additional chemical species equation to our
system:
\begin{equation} \label{biocideMu}
\mu(t,a,c) = \mu_0 \sigma(a) + \mu_1(t,c) (1-\sigma(a)).
\end{equation}
This form assumes that older cells are more resistant to antimicrobial challenge than
younger cells, and that the antimicrobial agent affects metabolically active cells
more than less active cells, represented by the oxygen dependence of
$\mu_1$.  In particular, we take
\begin{equation} \label{biocideMuForm}
 \mu_1(t,c) = \left\{
     \begin{array}{rl}
       \frac{50 c}{k+c}, & \qquad 35 \leq t \leq 35.2, \\
       0, & \qquad \mbox{otherwise.}
     \end{array}
   \right.
\end{equation}

\begin{figure}
\begin{center}
\includegraphics[height=1.6in]{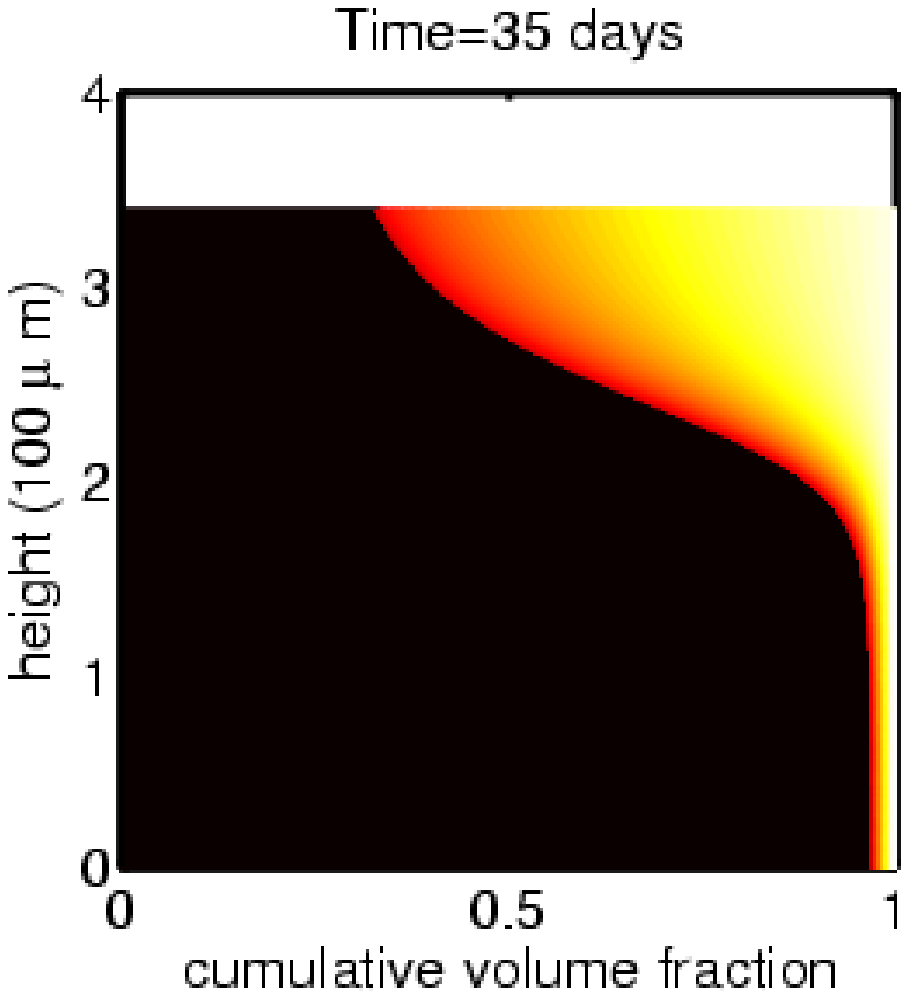}
\includegraphics[height=1.6in]{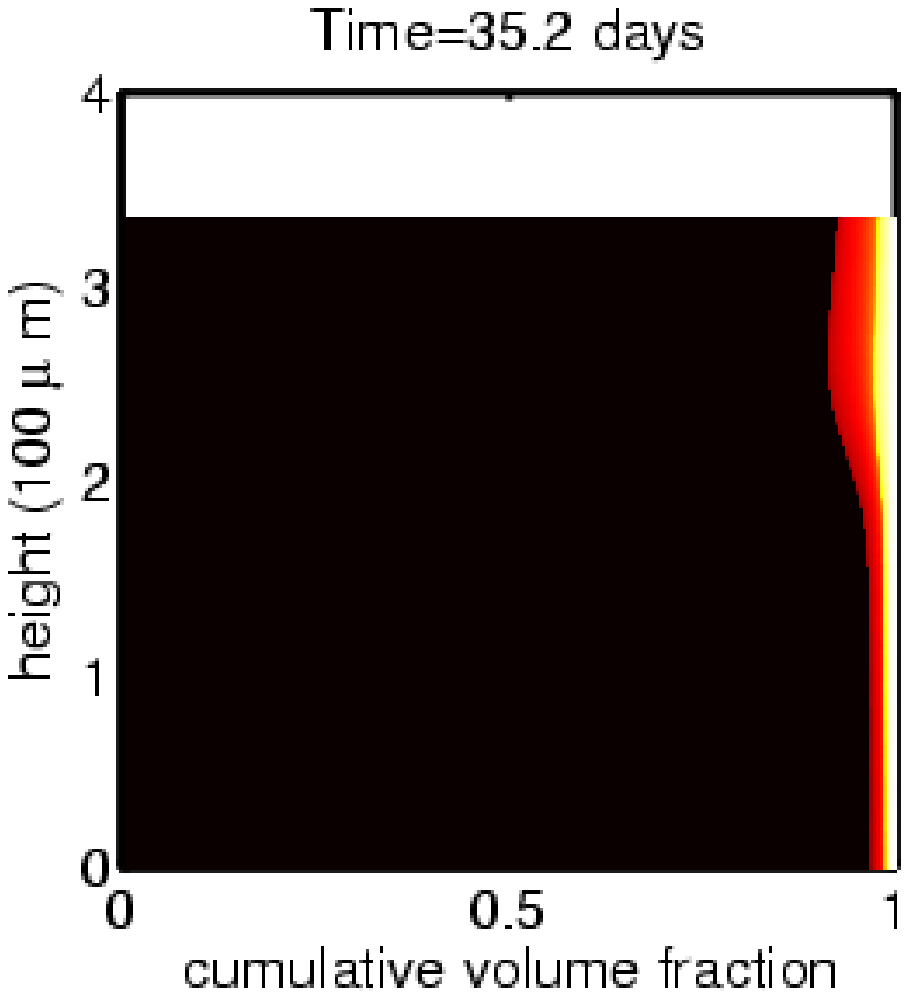}
\includegraphics[height=1.6in]{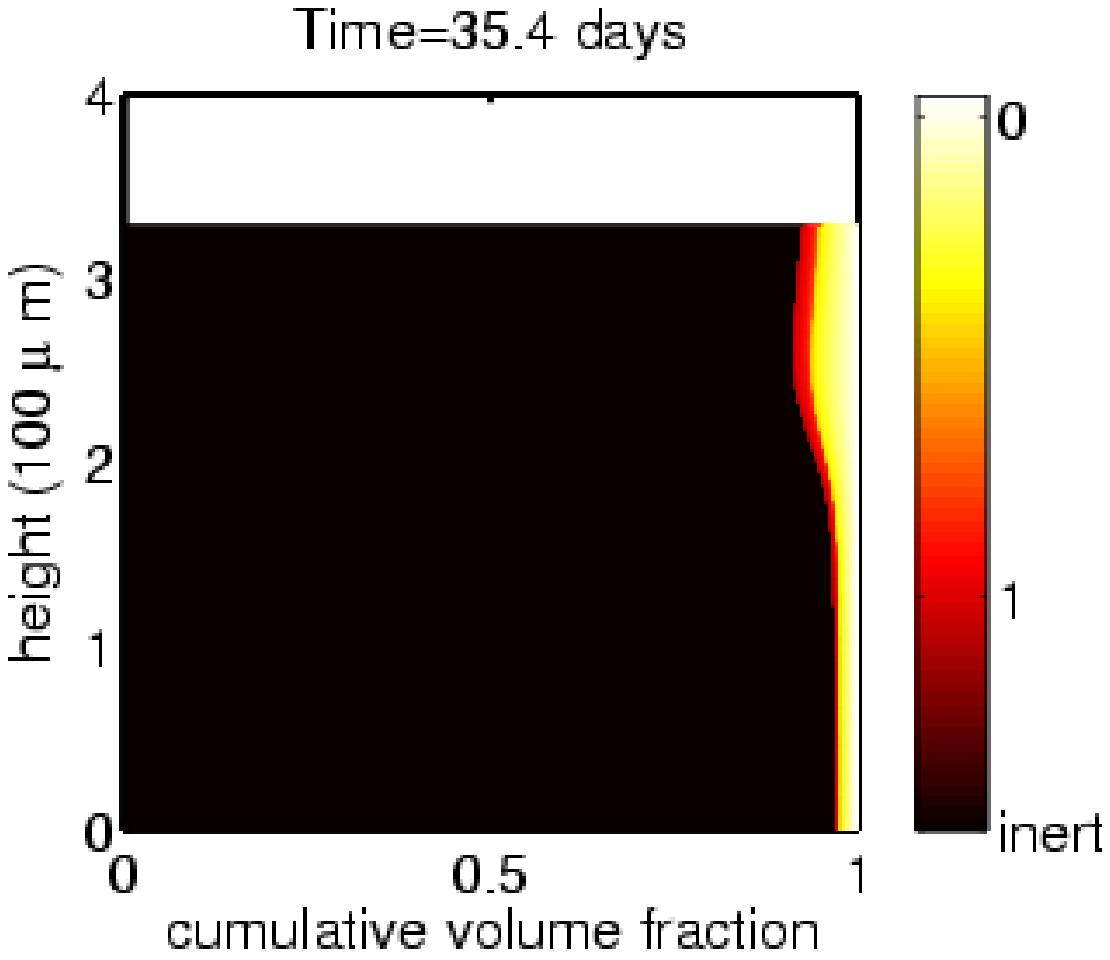} \\ 
\hspace{-0.24in} 
\includegraphics[height=1.6in]{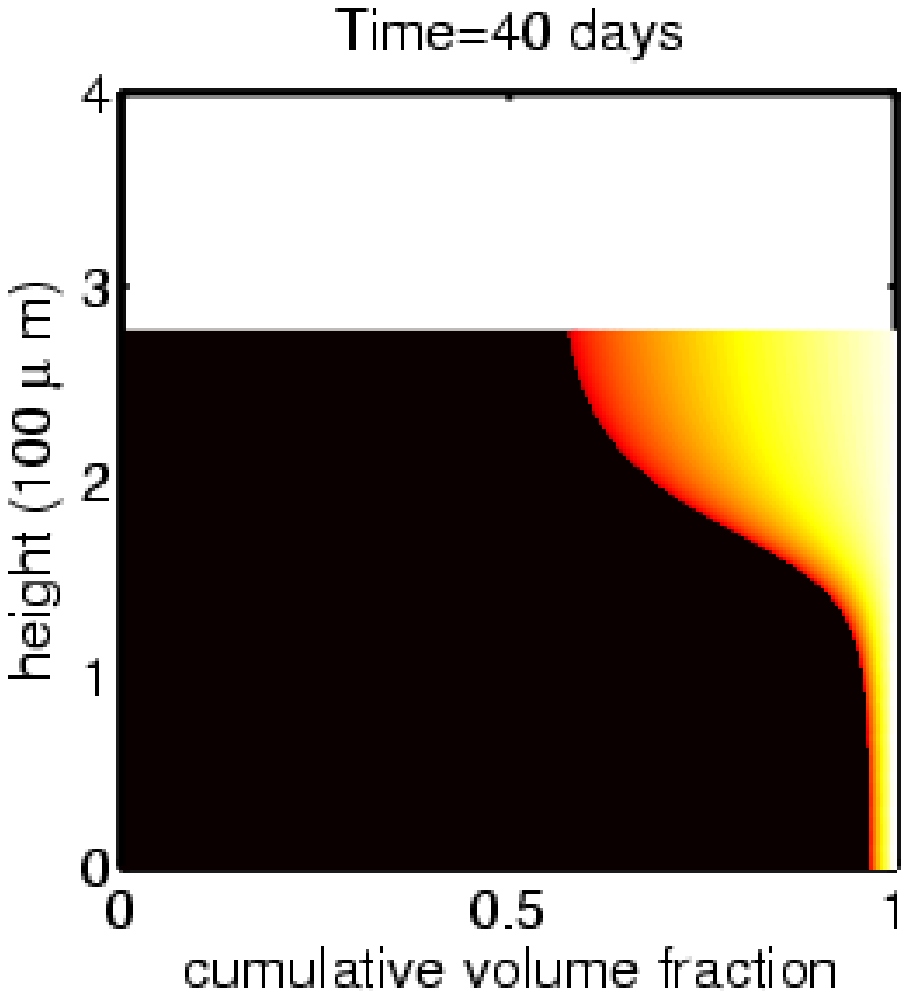}
\includegraphics[height=1.6in]{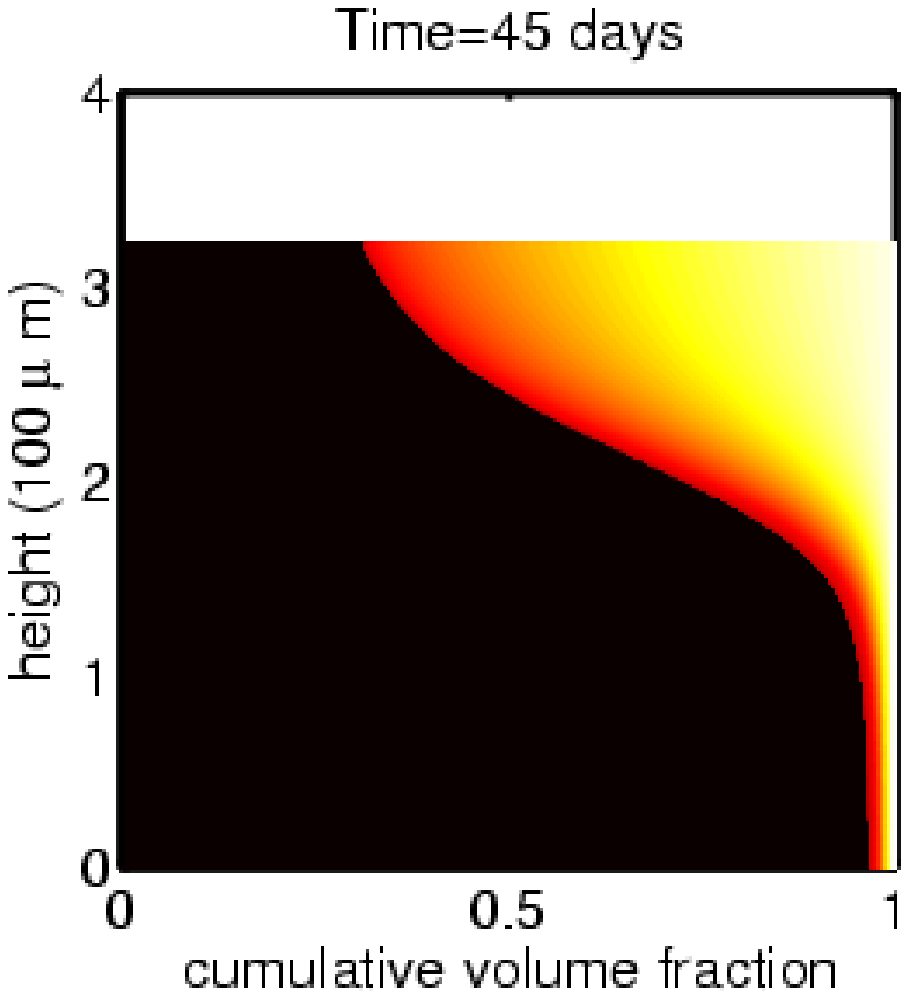}
\includegraphics[height=1.6in]{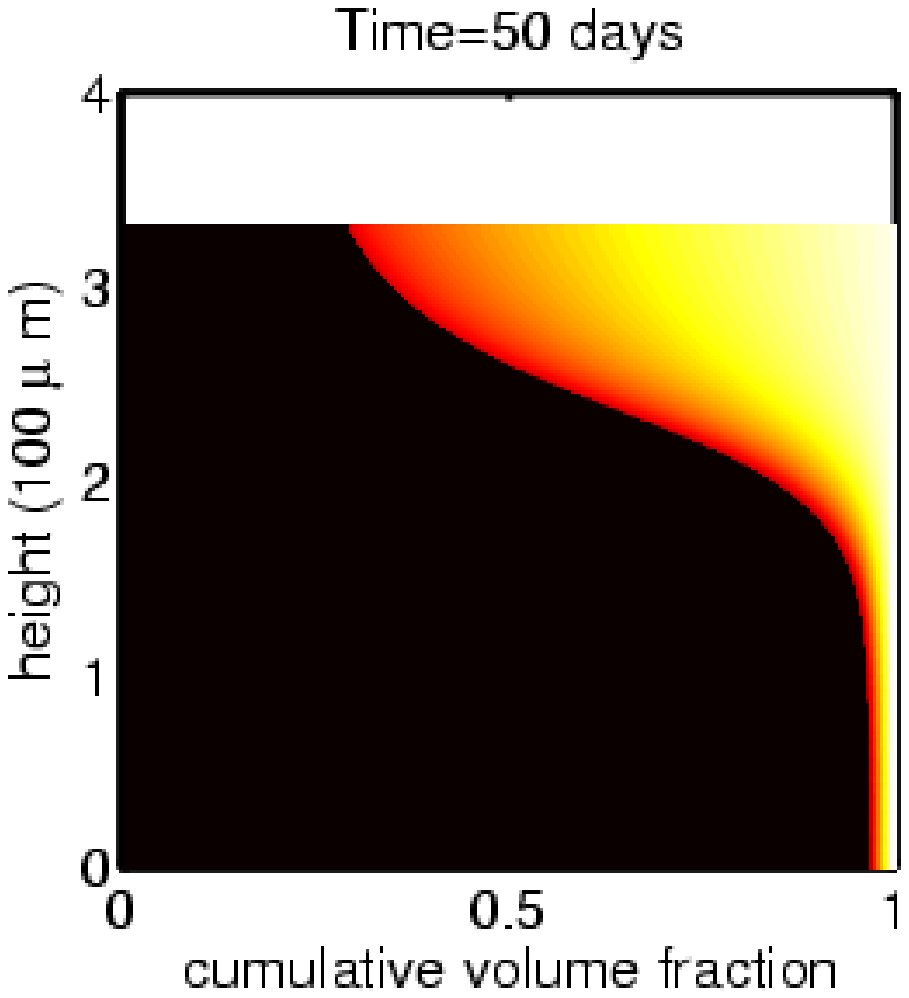}

\caption{Response to an antimicrobial agent applied from time $t=35$ to time $t=35.2$. The height of the colored area represents
the height of the biofilm, including both active and inert bacteria.
Color represents cell state: black represents inert cells, and a spectrum from
off-white to yellow to orange to red represents senescence of a cell
of a given age, $\sigma(a)$.  The horizontal width of a color
constitutes the volume fraction of cells of the corresponding senescence.}
\label{biocide}
\end{center}
\end{figure}

Results are shown in Figure \ref{biocide} for the case when an antimicrobial agent is applied from time $t=35$ to time $t=35.2$.   The senescence structure
of the population in the stalk allows it to maintain itself even after
the active layer is largely decimated.   Moreover, upon removal of the
antimicrobial agent, the population of older cells in the active layer
is quickly replaced by younger cells, which then return
the biofilm to its steady state over a longer maturation time. Note that the height continues to drop
after removal of the antimicrobial agent prior to regrowth.

\subsection{Numerical Methods}
\label{numerics}

We employ a moving-grid Galerkin method in age,
using piecewise constants as the approximation space
\cite{age-pwconst-paper}.  The use of higher-order approximation
spaces in age was discussed in \cite{age-general-paper}.  The
moving-grid Galerkin method decouples the age and time
discretizations, while allowing age and time to advance together along
characteristic lines.  Consequently, we are able to solve the model
equations in age without numerical dispersal or oscillations.  Because
the transport in age is computed by the movement of the grid, rather
than by a difference approximation of the age derivative or through
jump terms in a standard discontinuous Galerkin method, the only
meaningful source of error is approximation error, which underlies the
superconvergence results in \cite{age-pwconst-paper,age-general-paper}.  For the case of
piecewise constant functions, we obtain a second-order correct method in
age.

We integrate time using a step-doubling method
\cite{step-doubling-paper}.  Step-doubling consists
of taking one step of backward Euler over a time step, and then taking
two half steps of backward Euler over the same time interval.  This
results in two things.   First, we can compare the two late-time
solutions for the error control needed for the adaptivity in time.
Second, we can extrapolate the two solutions to get a likely
second-order accurate solution in time.

For the spatial variable, we discretize, over a uniform partition, the domain
$[0,\Gamma_t]$ and compute the changes in biofilm height by solving
equation (\ref{Gamma_t}).  We impose a boundary condition on $c$ at $\Gamma_t$ by using a ghost
node positioned at $\Gamma_{H_b}$.  Fluxes are computed using upwind differencing \cite{LeVeque2002}.
Although this method is only first-order correct, it has been
sufficient for the computations presented in this paper, given the
lack of sharp fronts in the interior of the biofilm, $[0,\Gamma_t]$.
More advanced methods will be needed for computations with more spatial dimensions.

The discretizations in the computational results presented above used a uniform partition of the spatial interval
$[0,\Gamma_t]$ with 301 nodes, and a
uniform age discretization of the truncated age domain, $[0,16]$,
with $\Delta a = 1/8$ and piecewise
constant basis functions.  A uniform age discretization in the context
of the moving-grid Galerkin
method means that all but the first and last age intervals are
constant in length, and that a new age interval is introduced at the
birth boundary when the old birth interval reaches $\Delta a$ in
length.
The tolerance parameter for the adaptive time-stepping in the
step-doubling algorithm was $5\times 10^{-3}$.

\section{Conclusions}
\label{conclusions}

In this paper we presented a multiscale model and simulation of
biofilm development that is interesting for three major reasons.  One is
the nonstandard multiscale nature of the problem: cell division and
aging is a result of complex, and fast, micro- and nano-scale
processes, at least when compared to the advective scale of the
biofilm growth.   However, by representing the cell division and aging process
using notions of senescence and age, we have a mechanism for the cellular
scale that, in keeping with what has been observed in other age- and
space-structured multiscale systems
\cite{JMB-proteus,MMStumor06,EnS,FnPnW}, is in general slower than
the advective process.   But even here we see novelty; unlike
\cite{JMB-proteus,MMStumor06,EnS,FnPnW},
the relative ranking of the time scales of the aging and advective
processes inverts as we move from an active layer at the top of the
biofilm to a passive layer below. Further, both of these times scales
are fast with respect to the biofilm maturation time.

This inversion of the time scales underlies another major point
of interest: the implication that the active layer does
not merely provide a physical shield for a reservoir of cells in the
passive layer, but also induces the passive layer to consist of an
increased proportion of senescent persister cells.

A third point of interest, and one which may have relevance to other
biological systems that exist in a polymer matrix, e.g. tumor-matrix
interactions, is the novel inclusion of age structure in a spatial
model where movement is due to growth-driven expansive stress rather
than diffusion or diffusion-like terms that represent mechanisms such
as chemotaxis or haptotaxis (movement of cells up a matrix gradient).

The model in this paper has a number of entailments  for future work.
One is experimental verification of the hypothesis that passive layers
in biofilm contain a disproportionate number of persister cells.
Another is a generalization of the model to higher spatial dimensions, and a study to
see in what manner the physical stalk of the mushroom-like shapes biofilm
often form affects the persister ``stalk'' visualized in the senescence
structure in this paper.   Finally, it is likely that many of the
modeling and simulation ideas developed in this paper have relevance
to other systems.  For example, inclusion of growth-driven expansive
stress into an age- and space-structured tumor model like that in \cite{MMStumor06}, alongside
other major mechanisms of motion such as diffusion and haptotaxis,
would result in models with more fidelity to the physical mechanisms of
tumor invasion.

\section*{Acknowledgments}
This work was initiated and much of it completed at IPAM during the Spring 2006
program {\em Cells and Materials}.   The authors thank the staff of IPAM for their
hospitality and support.

\bibliographystyle{siam}
\bibliography{cancer,age,na,math,bio,apps}

\end{document}